\documentclass[twocolumn]{aastex631}

\usepackage{amsmath}
\usepackage{epsf}
\usepackage{color}
\usepackage{enumitem}
\usepackage{mathtools}
\usepackage{float}
\usepackage[mathcal]{eucal}
 \let\mathscr\relax
\usepackage[scr]{rsfso}

\usepackage{pgffor}
\usepackage{makecell, tabularx}

\newcommand{\logM}{log$_{10}(M_\star/M_\odot)$}

\submitjournal{ApJ}

\graphicspath{{./}{figures/}}

\begin{document}

\title{Evidence for a Shallow Evolution in the Volume Densities of \\ Massive Galaxies at $z=4$ to $8$ from CEERS}

\author[0000-0003-4922-0613]{Katherine Chworowsky}\altaffiliation{NSF Graduate Fellow}
\affiliation{Department of Astronomy, The University of Texas at Austin, Austin, TX 78712 USA}

\author[0000-0001-8519-1130]{Steven L. Finkelstein}
\affiliation{Department of Astronomy, The University of Texas at Austin, Austin, TX 78712 USA}

\author[0000-0002-9604-343X]{Michael Boylan-Kolchin}
\affiliation{Department of Astronomy, The University of Texas at Austin, Austin, TX 78712 USA}

\author[0000-0001-8688-2443]{Elizabeth J.\ McGrath}
\affiliation{Department of Physics and Astronomy, Colby College, Waterville, ME 04901, USA}

\author[0000-0001-9298-3523]{Kartheik G. Iyer}
\affiliation{Dunlap Institute for Astronomy \& Astrophysics, University of Toronto, Toronto, ON M5S 3H4, Canada}

\author[0000-0001-7503-8482]{Casey Papovich}
\affiliation{Department of Physics and Astronomy, Texas A\&M University, College Station, TX, 77843-4242 USA}
\affiliation{George P.\ and Cynthia Woods Mitchell Institute for Fundamental Physics and Astronomy, Texas A\&M University, College Station, TX, 77843-4242 USA}

\author[0000-0001-5414-5131]{Mark Dickinson}
\affiliation{NSF's National Optical-Infrared Astronomy Research Laboratory, 950 N. Cherry Ave., Tucson, AZ 85719, USA}

\author[0000-0003-1282-7454]{Anthony J. Taylor}
\affiliation{Department of Astronomy, The University of Texas at Austin, Austin, TX 78712 USA}

\author[0000-0003-3466-035X]{{L. Y. Aaron} {Yung}}
\altaffiliation{NASA Postdoctoral Fellow}
\affiliation{Astrophysics Science Division, NASA Goddard Space Flight Center, 8800 Greenbelt Rd, Greenbelt, MD 20771, USA}
\affiliation{Space Telescope Science Institute, 3700 San Martin Dr., Baltimore, MD 21218, USA}

\author[0000-0002-7959-8783]{Pablo Arrabal Haro}
\affiliation{NSF's National Optical-Infrared Astronomy Research Laboratory, 950 N. Cherry Ave., Tucson, AZ 85719, USA}

\author[0000-0002-9921-9218]{Micaela B. Bagley}
\affiliation{Department of Astronomy, The University of Texas at Austin, Austin, TX 78712 USA}

\author[0000-0001-8534-7502]{Bren E. Backhaus}
\affiliation{Department of Physics, 196 Auditorium Road, Unit 3046, University of Connecticut, Storrs, CT 06269}

\author[0000-0003-0883-2226]{Rachana Bhatawdekar}
\affiliation{European Space Agency (ESA), European Space Astronomy Centre (ESAC), Camino Bajo del Castillo s/n, 28692 Villanueva de la Cañada, Madrid, Spain}

\author[0000-0001-8551-071X]{Yingjie Cheng}
\affiliation{University of Massachusetts Amherst, 710 North Pleasant Street, Amherst, MA 01003-9305, USA}

\author[0000-0001-7151-009X]{Nikko J. Cleri}
\affiliation{Department of Physics and Astronomy, Texas A\&M University, College Station, TX, 77843-4242 USA}
\affiliation{George P.\ and Cynthia Woods Mitchell Institute for Fundamental Physics and Astronomy, Texas A\&M University, College Station, TX, 77843-4242 USA}

\author[0000-0002-6348-1900]{Justin W. Cole}
\affiliation{Department of Physics and Astronomy, Texas A\&M University, College Station, TX, 77843-4242 USA}
\affiliation{George P.\ and Cynthia Woods Mitchell Institute for Fundamental Physics and Astronomy, Texas A\&M University, College Station, TX, 77843-4242 USA}

\author[0000-0003-1371-6019]{M. C. Cooper}
\affiliation{Department of Physics \& Astronomy, University of California, Irvine, 4129 Reines Hall, Irvine, CA 92697, USA}

\author[0000-0001-6820-0015]{Luca Costantin}
\affiliation{Centro de Astrobiolog\'ia (CAB), CSIC-INTA, Ctra de Ajalvir km 4, Torrej\'on de Ardoz, 28850, Madrid, Spain}

\author[0000-0003-4174-0374]{Avishai Dekel}
\affil{Racah Institute of Physics, The Hebrew University of Jerusalem, Jerusalem 91904, Israel}

\author[0000-0002-3560-8599]{Maximilien Franco}
\affiliation{Department of Astronomy, The University of Texas at Austin, Austin, TX, USA}

\author[0000-0001-7201-5066]{Seiji Fujimoto}
\affiliation{Cosmic Dawn Center (DAWN), Jagtvej 128, DK2200 Copenhagen N, Denmark}
\affiliation{Niels Bohr Institute, University of Copenhagen, Lyngbyvej 2, DK2100 Copenhagen \O, Denmark}

\author[0000-0003-4073-3236]{Christopher C. Hayward}
\affiliation{Center for Computational Astrophysics, Flatiron Institute, 162 Fifth Avenue, New York, NY 10010, USA}

\author[0000-0002-4884-6756]{Benne W. Holwerda}
\affil{Physics \& Astronomy Department, University of Louisville, 40292 KY, Louisville, USA}

\author[0000-0002-1416-8483]{Marc Huertas-Company}
\affil{Instituto de Astrof\'isica de Canarias, La Laguna, Tenerife, Spain}
\affil{Universidad de la Laguna, La Laguna, Tenerife, Spain}
\affil{Universit\'e Paris-Cit\'e, LERMA - Observatoire de Paris, PSL, Paris, France}

\author[0000-0002-3301-3321]{Michaela Hirschmann}
\affiliation{Institute of Physics, Laboratory of Galaxy Evolution, Ecole Polytechnique Fédérale de Lausanne (EPFL), Observatoire de Sauverny, 1290 Versoix, Switzerland}

\author[0000-0001-6251-4988]{Taylor A. Hutchison}
\altaffiliation{NASA Postdoctoral Fellow}
\affiliation{Astrophysics Science Division, NASA Goddard Space Flight Center, 8800 Greenbelt Rd, Greenbelt, MD 20771, USA}

\author[0000-0002-6610-2048]{Anton M. Koekemoer}
\affiliation{Space Telescope Science Institute, 3700 San Martin Dr., Baltimore, MD 21218, USA}

\author[0000-0003-2366-8858]{Rebecca L. Larson}
\affil{School of Physics and Astronomy, Rochester Institute of Technology, 84 Lomb Memorial Drive, Rochester, NY 14623, USA}

\author[0000-0001-7890-4964]{Zhaozhou Li}
\affil{Racah Institute of Physics, The Hebrew University of Jerusalem, Jerusalem 91904, Israel}

\author[0000-0002-7530-8857]{Arianna S. Long}
\altaffiliation{NASA Hubble Fellow}
\affiliation{Department of Astronomy, The University of Texas at Austin, Austin, TX, USA}

\author[0000-0003-1581-7825]{Ray A. Lucas}
\affiliation{Space Telescope Science Institute, 3700 San Martin Drive, Baltimore, MD 21218, USA}

\author[0000-0003-3382-5941]{Nor Pirzkal}
\affiliation{ESA/AURA Space Telescope Science Institute}

\author[0000-0002-9415-2296]{Giulia Rodighiero}
\affiliation{Department of Physics and Astronomy, Università degli Studi di Padova, Vicolo dell’Osservatorio 3, I-35122, Padova, Italy}
\affiliation{INAF - Osservatorio Astronomico di Padova, Vicolo dell’Osservatorio 5, I-35122, Padova, Italy}

\author[0000-0002-6748-6821]{Rachel S. Somerville}
\affiliation{Center for Computational Astrophysics, Flatiron Institute, 162 5th Avenue, New York, NY, 10010, USA}

\author[0000-0002-8163-0172]{Brittany N. Vanderhoof}
\affil{Rochester Institute of Technology, 84 Lomb Memorial Drive, Rochester, NY 14623, USA}

\author[0000-0002-6219-5558]{Alexander de la Vega}
\affiliation{Department of Physics and Astronomy, University of California, 900 University Ave, Riverside, CA 92521, USA}

\author[0000-0003-3903-6935]{Stephen M.~Wilkins} %
\affiliation{Astronomy Centre, University of Sussex, Falmer, Brighton BN1 9QH, UK}
\affiliation{Institute of Space Sciences and Astronomy, University of Malta, Msida MSD 2080, Malta}

\author[0000-0001-8835-7722]{Guang Yang}
\affiliation{Kapteyn Astronomical Institute, University of Groningen, P.O. Box 800, 9700 AV Groningen, The Netherlands}
\affiliation{SRON Netherlands Institute for Space Research, Postbus 800, 9700 AV Groningen, The Netherlands}

\author[0000-0002-7051-1100]{Jorge A. Zavala}
\affiliation{National Astronomical Observatory of Japan, 2-21-1 Osawa, Mitaka, Tokyo 181-8588, Japan}




\begin{abstract}

We analyze the evolution of massive (log$_{10}$ [$M_\star/M_\odot$] $>10$) galaxies at $z \sim$ 4--8 selected from the \textit{JWST} Cosmic Evolution Early Release Science (CEERS) survey. We infer the physical properties of all galaxies in the CEERS NIRCam imaging through spectral energy distribution (SED) fitting with \texttt{dense basis} to select a sample of high redshift massive galaxies. Where available we include constraints from additional CEERS observing modes, including 18 sources with MIRI photometric coverage, and 28 sources with spectroscopic confirmations from NIRSpec or NIRCam wide-field slitless spectroscopy. We sample the recovered posteriors in stellar mass from SED fitting to infer the volume densities of massive galaxies across cosmic time, taking into consideration the potential for sample contamination by active galactic nuclei (AGN). We find that the evolving abundance of massive galaxies tracks expectations based on a constant baryon conversion efficiency in dark matter halos for $z \sim$ 1--4.  At higher redshifts, we observe an excess abundance of massive galaxies relative to this simple model. These higher abundances can be explained by modest changes to star formation physics and/or the efficiencies with which star formation occurs in massive dark matter halos, and are not in tension with modern cosmology.

\end{abstract}


\section{Introduction} \label{sec:intro}

Measurements of galaxy stellar masses can provide a powerful benchmark to compare to theoretical models and simulations. The ability of the Universe to create and form galaxies with large amounts of stellar mass is largely dependent on galaxy baryonic feedback processes, both internally and within a galaxy's environment.
In particular, massive galaxies at high redshift provide interesting laboratories to study galaxy formation physics, as these extreme systems allow us to begin to understand how the Universe is able to rapidly build up large amounts of stellar mass in a short amount of time, and to constrain the feedback mechanisms that may be regulating these processes \citep[e.g.][]{Hayward2021}.

Observationally, extensive efforts have been made in attempting to understand how galaxy stellar masses evolve over time, with many studies measuring galaxy stellar mass functions (GSMFs) spanning cosmic time \citep[e.g.][]{Moffett2016, Weaver2022, Adams2021, McLeod2021MNRAS.503.4413M, Duncan2014, Muzzin2013,Song2016, Bhatawdekar2019, Tomczak2014, Wright2017, Kelvin2014, Stefanon2021, Leja2020}.
The stellar mass of a galaxy is typically inferred via modeling the rest-frame ultraviolet (UV) through near-infrared (NIR) spectral energy distribution (SED), which necessarily relies on assumptions about the galaxy's star formation history, metal enrichment history, dust attenuation, and stellar initial mass function (IMF).
Prior to \textit{JWST}, the \textit{Hubble Space Telescope} (\textit{HST}) provided the most accurate galaxy stellar mass estimates out to redshifts of $z \sim 5$, however, at increasingly high redshift, the wavelength coverage of \textit{HST} probes an increasingly limited wavelength range of the SED, and only UV-luminous galaxies are detected.
While the \textit{Spitzer} Infrared Array Camera (IRAC) provided longer wavelength space-based observations \citep[e.g.][]{Song2016, Bhatawdekar2019, Morishita2021, Tacchella2022}, the poor spatial resolution ($\sim 2"$) and small collecting area of \textit{Spitzer} IRAC with respect to \textit{HST} meant that galaxy observations were often blended with neighboring sources, limiting the ability of observstions to accurately attribute observed fluxes to their sources. 

The launch of \textit{JWST} \citep{JWSTmission} enabled a huge leap forward in both the resolution and sensitivity of infrared imaging.
With the longer wavelength coverage of \textit{JWST} Near Infrared Camera (NIRcam; \citealt{NIRCam_instrument}), observations of the rest-frame optical (including the star-formation history sensitive 4000 \AA\ break) are accessible out to $z\sim 10$, allowing for the measurement of accurate stellar masses out to redshifts that were previously inaccessible. Nearly two years since the launch of \textit{JWST}, data collected has already begun to revolutionize our understanding of the Universe, with new discoveries answering long-standing questions in the field of extra-galactic science, while also providing unexpected results and generating new questions. 

One of these unexpected early results is an apparent excess of UV-luminous $z \gtrsim 8$ galaxies relative to many recent theoretical models \citep[e.g.][]{Castellano2022,Naidu2022,Finkelstein2022,Finkelstein2023,Finkelstein2023b, Casey2023, Franco2023, Leung2023, Adams2023arXiv230413721A, Bouwens2023, Harikane2023}. 
This population of galaxies may indicate evidence for different star formation and stellar feedback physics in the early Universe, leading to earlier, more rapid formation of massive and luminous galaxies than previously thought possible \citep[e.g.][]{Mason2023, Shen2023, Pallottini2023, Yung2023, Dekel2023, Ferrara2023}. While models shown tension with direct observables, these UV luminosities are very sensitive to assumed initial mass functions, dust, and AGN contamination. Having reliable estimates for observed stellar masses provides an alternative way of testing this putative tension with models.
Excitingly, early {\it JWST} studies have claimed evidence for very massive galaxies (some exceeding \logM $>$ 11), further indicating the possibility of more efficient star-formation activity at higher redshift than expected \citep[e.g.][]{Labbe2023, xiao2023massive}, and/or a revision to $\Lambda$CDM \citep{MBK2023}. 

To improve our understanding of the abundance of massive (defined here as log$_{10}[M_\star/M_\odot]$ $>10$) galaxies and how this population evolves over cosmic time, we present a study of massive galaxies selected over the full the 88 arcmin$^2$ Comic Evolution Early Release Survey (CEERS) NIRCam field. We briefly introduce CEERS and the data reduction (\S \ref{sec:data}) before discussing how the galaxies studied here were selected (\S \ref{sec:Selection}), focusing on sources at $z>4$. In selecting massive galaxies, we also note the interesting prominence of a population of  galaxies arising from recent \textit{JWST} observations: point-like sources with flat or blue rest UV-optical spectra alongside increasing brightness at longer wavelength, possibly hosting dust-reddened active galactic nuclei \citep[AGN][]{Barro2023, Kocevski2023, Greene2023, Labbe2023UNCOVER}. We discuss how this category of sources may bias selection of massive galaxies (\S \ref{subsec:AGN}). We present how the cumulative number density of massive galaxies evolves across cosmic time in Section \ref{sec:results} and compare our findings to both previous studies (\S \ref{subsec:comp_obs}) and model predictions (\S \ref{subsec:comp_models}). Finally, we discuss the our observed abundances of galaxies may indicate changes to the efficiency of star-formation in galaxies (\S \ref{subsubsec:SFefficiency}) or how the observed light from the galaxy is translated to mass (\S \ref{subsubsec:MtoL}). Throughout this paper, we assume a cosmology of H$_0$=70 km s$^{-1}$ Mpc$^{-1}$, $\Omega_M$=0.3, $\Omega_\Lambda$ = 0.7. When necessary, we assume a baryonic density parameter of $\Omega_b h^2 = 0.02237$ \citep{Plack2016}. We use a \citet{ChabrierIMF} IMF, and all magnitudes given are in the AB system \citep{ABmag}.

\section{NIRCam \& HST Catalog} \label{sec:data}
We perform our analysis on the publicly released CEERS NIRCam mosaics (Data Releases 0.5 and 0.6), which include imaging from $1-5\micron$ in F115W, F150W, F200W, F277W, F356W, F410M and F444W and cover ten pointings in the Extended Groth Strip (EGS) \citep{bagley2023}.
The CEERS observations were completed and reduced over two epochs. Epoch 1 NIRCam pointings 1, 2, 3 and 6 were obtained in parallel to the Mid-Infrared Instrument (MIRI) imaging in June 2022. These pointings are reduced with \textit{JWST} Calibration Pipeline \citep{bushouse2023} version 1.7.2 and Calibration Reference Data System (CRDS) pmap 0989, and are available from CEERS DR0.5. The Epoch 2 observations released with CEERS DR0.6 include NIRCam imaging over pointings 4, 5, 7, 8, 9 and 10 obtained in December 2022 in parallel to NIRSpec MSA observations. Pointings 5, 7, 8, 9 were also observed with NIRCam WFSS with MIRI imaging in parallel. These Epoch 2 images were reduced with Pipeline version 1.8.5 and pmap 1023. The two CEERS NIRCam reductions follow identical procedures, with the only difference being in Pipeline and CRDS versions. 
We note that the only NIRCam update between pmaps 1089 and 1023 is to the distortion reference files, and so this change will not affect the noise properties, flat fields, or flux calibration between CEERS DR0.5 and DR0.6. 
The CEERS NIRCam imaging reduction uses the \textit{JWST} Calibration Pipeline, with custom modifications to correct for features and challenges in the data such as snowballs, wisps, 1/f noise and to improve the astrometric alignment.  We direct the reader to \citet{bagley2023} for further details on the reduction.

We make use of the photometric catalog from \citet{Finkelstein2023b}, which also includes the available {\it HST} imaging in this field from CEERS {\it HST} Data Release 1, including the ACS F606W, F814W and WFC3 F105W, F125W, F140W and F160W filters from {\it HST} programs including CANDELS \citep{grogin2011, koekemoer2011}.  This catalog's methodology is similar to \citet{Finkelstein2023}, with some updates to improve the accuracy of color and total flux measurements. 
Flux uncertainties were calculated empirically via measurements of the flux spread in randomly placed apertures.  We direct the reader to \citet{Finkelstein2023b} for full details on all cataloguing procedures.

\begin{figure*}[!t]
\centering
\includegraphics[width=0.9\linewidth]{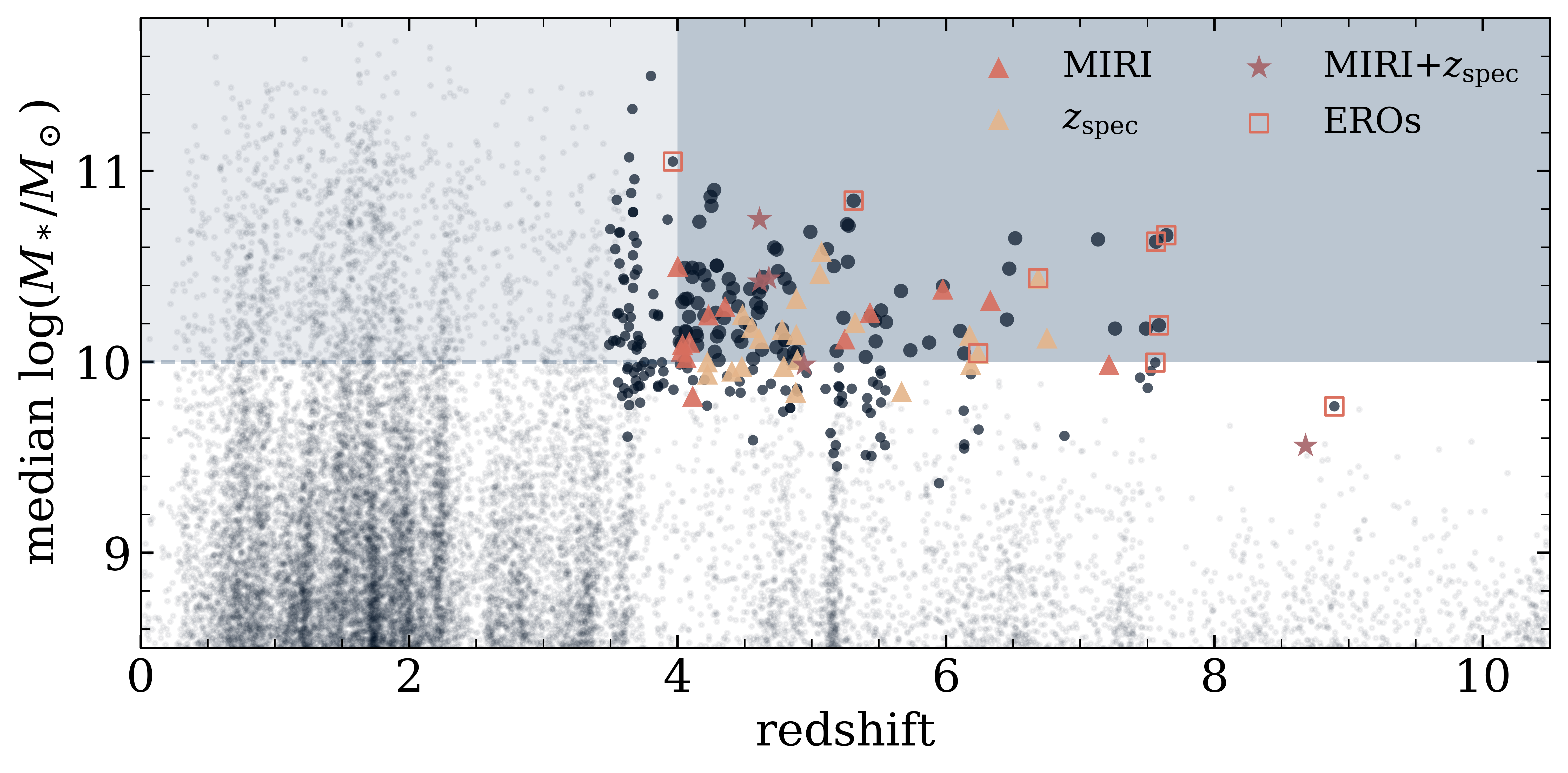}
\caption{The distribution in redshift and stellar mass of the full CEERS catalog \citep{Finkelstein2023b}, as measured by \texttt{dense basis} (with \texttt{eazy} priors on the \texttt{dense-basis}-derived redshift).  The darker shaded region denotes our redshift and mass range of interest, while the dark points indicate any source which has been visually inspected, the larger of which are sources whose median stellar mass and redshift satisfy our high redshift massive criteria. 
We also show galaxies with MIRI coverage (red triangles) and those with spectroscopic confirmation (orange triangles). The galaxies marked with a red box are ``Extremely Red Objects" (EROs): F444W point sources with red F277W-F444W colors that are likely hosts to dust-reddened AGN (discussion in \S \ref{subsec:AGN}). We find 118 galaxies with median stellar mass and redshift values in our region of interest, including objects up to $z \sim$7.5.  Notably, we find no sources at \logM\ $>$ 11 or at $z >$ 8, and the highest redshift sources we do find are all classified as EROs, thus reliable mass measurements are not possible.}
\label{fig:sample}
\end{figure*}

\section{Selection of Massive Galaxies} \label{sec:Selection}
Here, we detail our process of selecting a sample of massive galaxies. We first make use of the \texttt{eazy} \citep{Brammer2008} to determine photometric redshifts.  We then use the derived redshift probability distribution functions (PDFs, or $\mathcal{P}(z)$) as a prior to \texttt{dense basis} \citep{Iyer_2017, Iyer_2019}, which we use to perform Bayesian SED fitting to recover physical properties (i.e. stellar mass) and their corresponding posterior distributions.

\subsection{Photometric Redshift Fitting}\label{subsec:phot-z}
Photometric redshifts for the catalog were determined with \texttt{eazy}.  We use these results from \citet{Finkelstein2023b}, briefly discussing the procedure here.  \texttt{eazy} fits non-negative linear combinations of stellar population synthesis templates through a
user-defined grid of redshifts, and at each redshift finds
the best fitting synthetic template spectrum by minimizing $\chi^2$. \texttt{eazy} derives a $\mathcal{P}(z)$ based on the quality of fit to the observed photometry. The templates used include the “tweak fsps QSF 12 v3” set of 12 FSPS \citep{ConroyGunn2010} templates recommended by the \texttt{eazy} documentation as well as an additional six templates designed to encompass bluer rest-frame UV colors expected from $z>9$ galaxies \citep{Larson2022}.

\subsection{SED fitting}\label{subsec:SEDfitting}
We perform spectral energy distribution (SED) fitting for all sources with signal-to-noise in F444W $>3$ with the \texttt{dense basis} SED fitting code, which is designed to robustly recover star formation history (SFH) constraints from galaxy SEDs. \texttt{dense basis} uses a flexible non-parametric SFH represented by a Gaussian Mixture Model \citep[GMM; ][]{Iyer_2019}, and stellar templates generated from \texttt{FSPS} \citep{Conroy2009, ConroyGunn2010} including implementation of nebular emission lines using \texttt{CLOUDY} \citep{Ferland2017CLOUDY, Byler2017}.
For this work, we define three ``shape'' parameters that describe the SFH: $t_{25}, t_{50},$ and $t_{75}$ (requiring the recovered SFH of the galaxy to form ``$x$'' fraction of its total mass by time $t_x$). All sources were fit with \texttt{dense basis} assuming a \citet{Calzetti_2000} dust law and a Chabrier initial mass function \citep[IMF][]{ChabrierIMF}. We impose a uniform (flat) prior on the specific star formation rate (sSFR) with limits on the sSFR (sSFR$ \textrm{yr}^{-1} \in [-14, -7]$), an exponential prior on the dust attenuation over a wide range of values ($A_V \in [0, 4]$), and a uniform (in log-space) prior on the metallicity ($Z/Z_\odot \in [0.01, 2.0]$). 
For redshift, \texttt{dense basis} is only able to take a prior in the form of a top-hat function. Therefore, in order to input the photometric redshift information as recovered by \texttt{eazy}, we modify the \texttt{eazy} $\mathcal{P}(z)$ to a top-hat form by first fitting a Gaussian to the primary peak of the $\mathcal{P}(z)$, and taking the width of the tophat function as the $2\sigma$ width of the Gaussian, if $\sigma >1.5$, or the $5\sigma$ width of the Gaussian, if $\sigma \leq 1.5$ (in the case of a very narrowly peaked Gaussian). For each source, we thus have a redshift prior based on the \texttt{eazy} recovered redshift PDF.

While we perform SED fitting over the entire catalog and all available redshifts, as the emphasis of this work is on the evolution at $z >$ 4, we further vet that higher-redshift sample to ensure accurate results. Therefore, we split our massive sample into two categories: 1. a low redshift sample of 1796 sources with best fit redshift $z < 3.5$ and recovered median stellar masses of \logM $>10$, and 2. a parent high redshift sample, with a photometric redshift of $z>3.5$ (from \texttt{dense basis}), at least 70\% of the integrated \texttt{eazy} $\mathcal{P}(z)$ greater than 2.5, and a less stringent mass cut of 2.5\% of the posterior in stellar mass having \logM$> 10$. The parent high redshift sample is intentionally chosen to include sources which have median stellar masses below our intended mass limit, to ensure we perform visual inspection on all sources where posteriors may include a massive (\logM$>10$) solution to assess the reliability of the photometry. We thus manually inspect 561 sources satisfying these criteria.

We remove a total of 297 sources from our sample due to unreliable photometry (e.g. too close to a detector edge, bad pixels, spurious sources from \texttt{Source Extractor} \citep{Bertin2010SExtractor}, and too close to a neighboring bright source), the majority (168/297) sources removed have $z_{\mathrm{phot}} > 8$, therefore we are confident in their spurious nature.
Two particularly high redshift sources were determined to be incorrectly fit at high redshift (IDs 52092 and 75728 at $z_{\textrm{phot}}\sim 15$ and $z_{\textrm{phot}} \sim 12$, respectively). Source ID 52092 has a low-$z$ solution with $\Delta \chi^2 < 4$ from the high-$z$ fit, and source ID 75728 had significant flux detected below the putative Lyman-$\alpha$ break at its fit redshift. Both sources were thus refit with \texttt{dense basis} at the low redshift solution from \texttt{eazy}. After refitting, one source was removed from the sample for not satisfying our mass limit. 

53 additional sources were ``oversplit", where multiple sources in the photometric catalog appear, based on visual inspection, to be part of a single large galaxy (this is because the \texttt{Source Extractor} run  was optimized to find high redshift, faint sources). For these galaxies, we further inspect the segmentation map to determine the neighboring components that are likely to be from the same galaxy. The photometry of each component was combined if the individual best-fitting redshift was within $\Delta z \leq 0.5$. The errors on flux are added in quadrature and the updated photometry was then refitted both with \texttt{eazy} and \texttt{dense basis} as described above. After correcting the photometry, 50/53 sources were still considered high redshift and massive.

This resulted in 261 sources with robust photometry which satisfied our selection. While the mass limit in this initial selection implies that many of these galaxies in fact have median stellar masses estimated to be below our intended mass limit, we draw stellar mass posteriors from this larger sample when measuring the volume densities of massive galaxies (details in \S \ref{subsec:number_densities}), to account for the tail of their mass posterior which exceeds our mass limit of \logM\ $>$ 10.
Figure \ref{fig:sample} shows the redshift and stellar mass distribution recovered here. 


While we expect there to exist similar contaminants and spurious sources in the lower redshift sample, we elect not to inspect the complete sample as this would be unfeasible for such a high number of sources, we do however inspect a random subsample of the low redshift sample and found a very low spurious rate, therefore the expected contamination is not expected contribute significantly to the derived statistics from the sample due to the large number of real sources. Therefore the following discussion primarily focuses on the visually inspected high redshift subsample of 261 massive galaxies presented here.

\begin{figure*}[t!]
\centering
\includegraphics[width=0.9\linewidth]{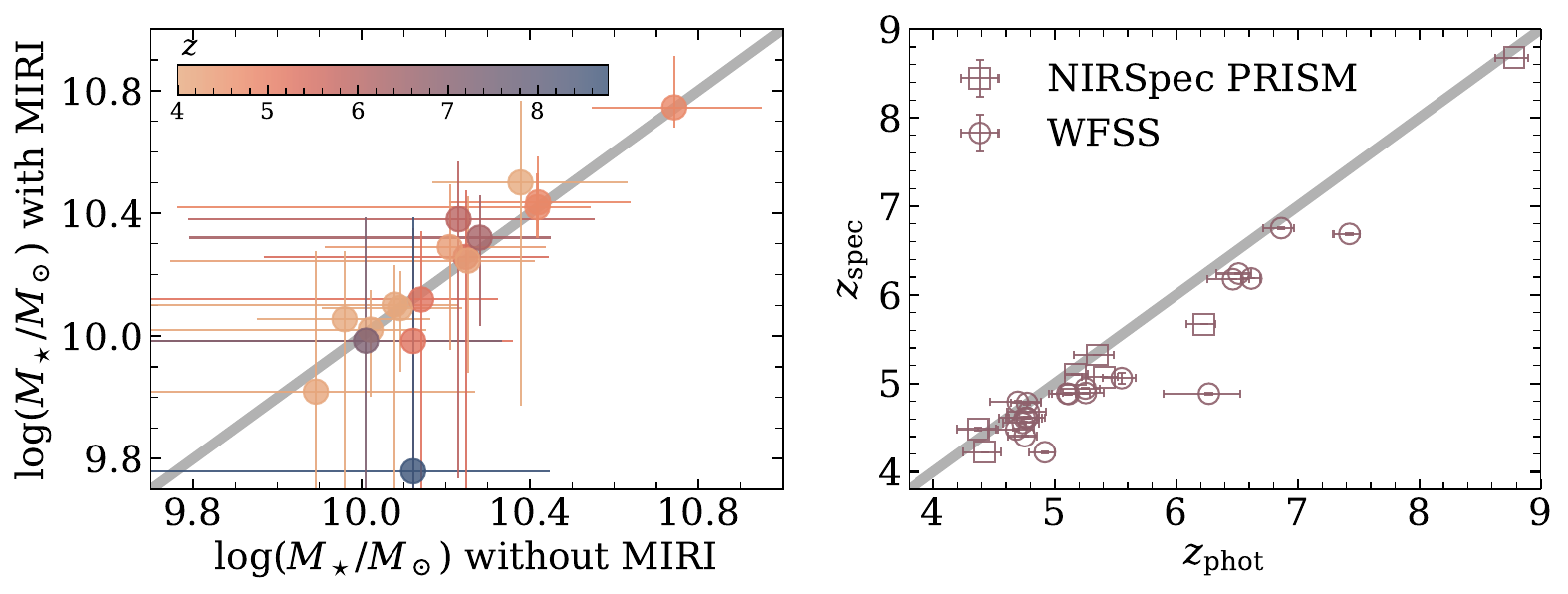}
\caption{\textit{Left:} Estimated stellar masses for galaxies with and without MIRI photometry. With one exception (the \citet{Larson2023} galaxy, which also has spectroscopic confirmation), including MIRI photometric constraints does not significantly alter the recovered stellar masses (average $\Delta M_\star < \sim 0.1$ dex), though it does reduce the median error by $\sim 0.05$ dex. \textit{Right:} Comparison of photometric to measured spectroscopic redshifts for sources covered by NIRSpec (five sources) and WFSS (22 sources). Similar to previous analyses of \textit{JWST} spectroscopy \citep[e.g.][]{Fujimoto2023, ArrabalHaro2023}, we do observe a slight systematic offset between photometric redshift estimates and confirmed spectroscopic redshifts, with all but two galaxies having confirmed spectroscopic redshifts lower than the best fit photometric estimate. However, we find general good agreement between $z_\textrm{phot}$ and $z_\textrm{spec}$, with only four sources with $\Delta z > 0.5$, and no catastrophic ouliers.  We update our SED fitting with spectroscopic redshifts when available. \label{fig:other_obs}}
\end{figure*}

\subsection{Additional Data} \label{subsec:AdditionalData}
The CEERS field includes coverage by both MIRI imaging \citep{MIRI}, NIRSpec spectroscopy \citep{NIRSpec}, and NIRCam Wide field slitless spectroscopy (WFSS; \citealt{NIRCam_instrument}). NIRSpec allows for detection of  strong [O {\sc iii}] emission out to $z<9.6$, while the CEERS WFSS observations with the F356W filter are sensitive to [O {\sc iii}] redshifts at $5 < z<7$, allowing spectroscopic confirmation for sources with strong emission lines in our sample. 

\subsubsection{MIRI} \label{subsubsec:MIRI}
Eighteen high redshift massive galaxies fall within the area covered by the CEERS MIRI imaging, extending the wavelength coverage for SED fitting to longer wavelengths.
The MIRI imaging is conducted over eight pointings, four blue pointings (P3, P6, P7, P9) providing deep imaging with F560W and F770W, and four with contiguous wavelength coverage in F1000W, F1280W, F1500W, and F1800W (P1, P2, P5, P8). Two of these pointings (P1, P2) also include coverage in F770W and F2100W. These observations were performed in two epochs, in June 2022 and December 2022. Pointings 1, 2, 3, and 6 were observed in Epoch 1 as prime observations with parallel NIRCam imaging, and pointings 5, 7, 8, and 9 were observed in parallel to NIRCam Wide Field Slitless Spectroscopy. Details of the MIRI data reduction can be found in \citet{yang2023ceers}.

\texttt{Source Extractor} v2.19.5 \citep{SourceExtractor} was run on the blue pointings using F560W+F770W as detection images, and using the PSF-matched F560W and F770W images and their associated RMS maps for flux measurements \citep{Papovich2023}.
In the red MIRI pointings, photometry extraction was determined using \texttt{TPhot} v2.1 \citep{Tphot2} following \citet{Yang2021}. \texttt{TPhot} was used for the red pointings instead of \texttt{Source Extractor} due to the large wavelength coverage of these pointings and to mitigate quality loss by PSF matching across filters. 
Of the galaxies presented here, 18 were covered in MIRI (16 in the deeper blue pointings, and 2 in the red pointings). We perform our SED fitting including MIRI constraints and re-derive their physical parameters. We find that the stellar masses largely agree between SED fitting with and without MIRI observations within errors, however, including MIRI photometry slightly decreases the stellar masses of the sample by $\Delta M_\star \sim 0.02$ dex (Figure \ref{fig:other_obs}), with 1/18 source (the \citealt{Larson2023} galaxy) having a decrease in measured stellar mass of 0.5 dex, similar to results found by \citet{Papovich2023}. The
inclusion of these longer wavelength observations does provide marginally tighter constraints on the estimated stellar masses, reducing the median error bar by $\sim 0.05$ dex. 

\begin{figure*}[!t]
\centering
\includegraphics[width=0.95\linewidth]{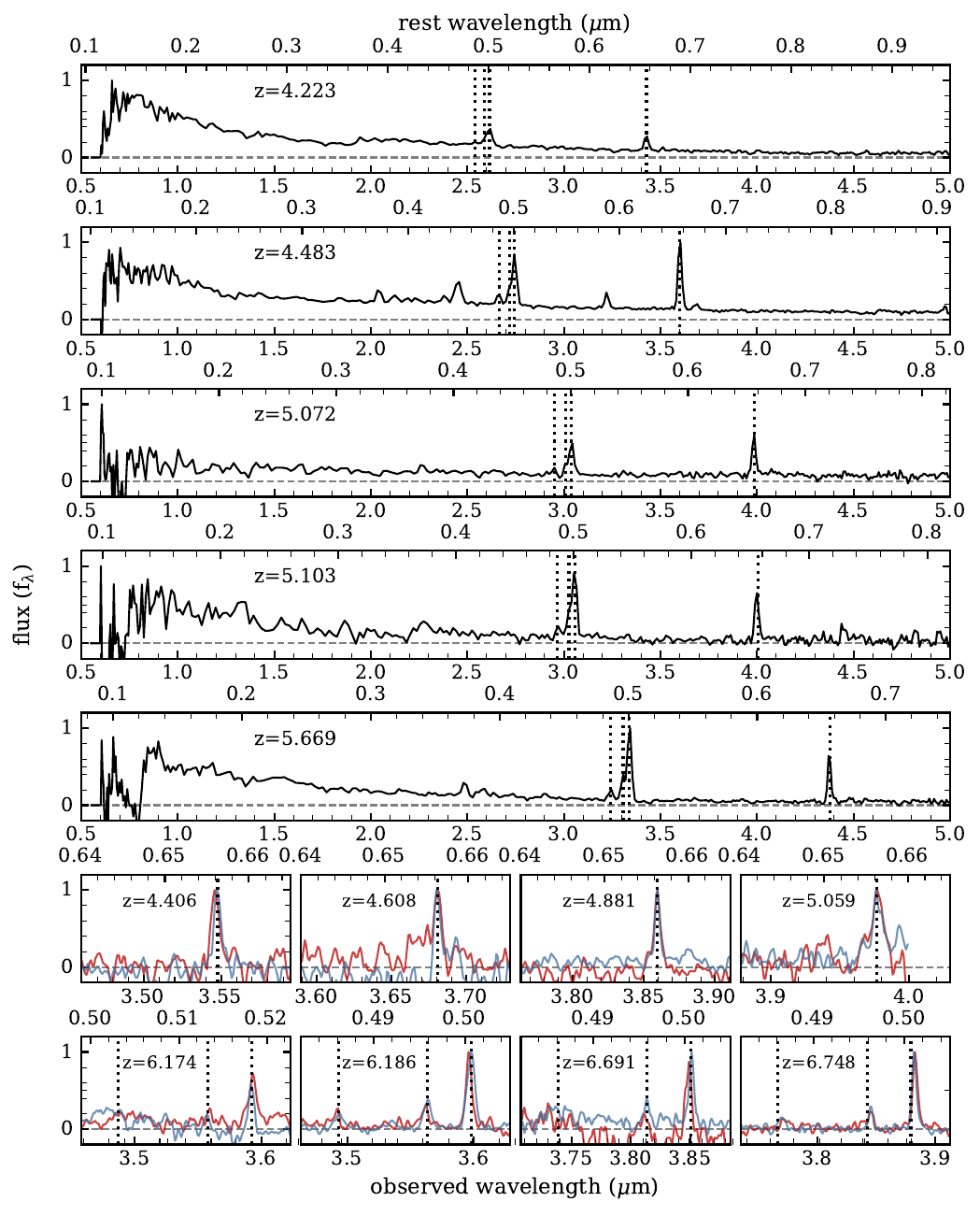}
\caption{Examples of spectroscopic observations of sources within our sample of \logM $>10$ galaxies. The top five panels show spectra for the five sources with NIRSpec PRISM observations (one additional source, presented in \citealt{Larson2023}, has a NIRSpec grating redshift). Each galaxy shows clearly detected [O{\sc iii}] doublet+H$\beta$ and H$\alpha$ emission. The bottom eight panels show line detections in the GRISM, with the top row showing H$\alpha$ and the bottom row shows the [O{\sc iii}] doublet+H$\beta$. The red (blue) shows the row (column)-dispersion. Here we smooth the GRISM data using a Gaussian filter with a width of one spectral element for clarity. 
28 sources have clear emission lines providing robust spectroscopic redshift confirmation of many of the sources in our sample.}
\label{fig:sample_spec}
\end{figure*}

\subsubsection{Spectroscopy} \label{subsubsec:Spectroscopy}
For this sample, five sources were observed with NIRSpec with the prism, and one object was also observed with the medium-resolution gratings.  The data processing is explained in detail in Arrabal Haro et al. (in prep.). However, the main steps of the reduction follow those employed in \citet{ArrabalHaro2023, Fujimoto2023, Kocevski2023, Larson2023}, and is briefly summarized here. These NIRSpec data were reduced using \textit{JWST} Calibration Pipeline v1.8.5 and CRDS mapping 1045.
Custom parameters were utilized to improve correction on ``snowball" events. The resulting count-rate maps are processed through stage two using \verb|calwebb_spec2| pipeline modules,
the images of the three nods are combined to extract the one-dimensional spectra.

For all six sources, strong emission lines were observed, including [O{\sc iii}] and $H\alpha$, providing clear spectroscopic redshift confirmations. One of these sources was previously presented in \citet{Larson2023}, thus we use their measurements (the source with medium-resolution grating observations).  For the remaining five we measure spectroscopic redshifts by fitting Gaussian functions to observed lines, converting to redshifts using the relevant vacuum rest-frame wavelength. 

Eighty-two sources were covered by F356W WFSS observations. Of these 82 sources, 22 have clearly identifiable emission lines in both the row and column dispersions. The WFSS spectra were extracted following the Simulation Based Extraction method which was described in \citet{Pirzkal17}.  To summarize, we used the CEERS master mosaic and available photometry to duly simulate each individual WFSS observations and these served to estimate and correct for spectral contamination in the 2D WFSS wavelength calibrated spectra, before these were combined and used to produce a pair of independent  (GRISMR and GRISMC) 1D spectra using optimal extraction. We used the latest, most up-to-date version of the NIRCAM WFSS calibration products which were finalized in August 2023. The estimated accuracy of the field dependence of the calibration is 0.2 pixel while the estimated uncertainty in the final wavelength calibration is 2.5\AA{}.
We inspected each of the objects by eye and identified objects with obvious emission lines present in both the row and column dispersion data. We conservatively require detections in both observations to ensure no spurious lines were included; several sources contain partial observations in only one dispersion direction, thus future work may increase the yield of spectroscopic observations.  We used a combination of the photometric redshifts and the presence or lack of a clear line doublet to classify the strong emission lines as H$\alpha$ or [O{\sc iii}]$\lambda\lambda$5007,4959 lines. For the H$\alpha$ objects, we fit a single Gaussian function with a local continuum offset to the emission line, and for the [O{\sc iii}] objects we fit two Gaussians with a fixed 2.98:1 amplitude ratio, a fixed line-center ratio of 5006.843:4958.911, a common Gaussian width, and a local continuum offset. Using a Monte Carlo technique, we fit the appropriate model to each object 1000 times, and varied the flux data of the object spectrum using the spectral error data between each run. We use the median line centers and their standard deviations from these Monte Carlo runs to determine spectroscopic redshifts with uncertainties for each object. We report these values in Table \ref{tab:galaxy_properties}.

\begin{deluxetable*}{lcccccccc}
\vspace{2mm}
\tablecaption{Properties of Massive Galaxy Sample}
\tablewidth{\textwidth}
\tablehead{
\multicolumn{1}{c}{ID} &
\multicolumn{1}{c}{RA} &
\multicolumn{1}{c}{Dec} &
\multicolumn{1}{c}{Flux F277W} &
\multicolumn{1}{c}{Stellar Mass} & 
\multicolumn{1}{c}{$z_{\mathrm{phot}}$} & 
\multicolumn{1}{c}{$z_{\mathrm{spec}}$} & 
\multicolumn{1}{c}{MIRI} \\ &&& $M_{\mathrm{AB}}$ & \multicolumn{1}{c}{\logM} }
\startdata
\hline
CEERS-42006      & 214.73721 & 52.72742 &  22.9 & $ 10.90^{+0.25}_{-0.33}$ & $4.27^{+0.25}_{-0.23}$ & --& -- \\
CEERS-93955      & 214.77138 & 52.74975 &  25.0 & $ 10.86^{+0.32}_{-0.19}$ & $4.25^{+0.50}_{-0.48}$ & --& -- \\
CEERS-37979$^*$  & 214.79537 & 52.78885 &  24.7 & $ 10.84^{+0.10}_{-0.60}$ & $5.31^{+0.10}_{-0.10}$ & --& -- \\
CEERS-45649      & 214.89649 & 52.87087 &  22.9 & $ 10.82^{+0.16}_{-0.50}$ & $4.25^{+0.16}_{-0.14}$ & --& -- \\
CEERS-80676      & 215.04413 & 52.89874 &  23.1 & $ 10.75^{+0.17}_{-0.07}$ & $4.77^{+0.10}_{-0.17}$ & 4.611$^\ddag$ & yes\\
CEERS-8028       & 215.03905 & 53.00278 &  23.9 & $ 10.73^{+0.24}_{-0.19}$ & $4.16^{+0.26}_{-0.30}$ &  --& -- \\
CEERS-10933      & 214.90311 & 52.94573 &  23.5 & $ 10.72^{+0.24}_{-0.43}$ & $5.26^{+0.15}_{-0.17}$ &  --& -- \\
CEERS-2167       & 215.01122 & 53.01374 &  23.6 & $ 10.71^{+0.29}_{-0.78}$ & $5.27^{+0.13}_{-0.17}$ &  --& -- \\
CEERS-13082      & 214.91381 & 52.94294 &  23.6 & $ 10.68^{+0.20}_{-0.38}$ & $4.99^{+0.13}_{-0.17}$ &  --& -- \\
CEERS-9317$^*$   & 214.98304 & 52.95601 &  25.7 & $ 10.66^{+0.25}_{-0.22}$ & $7.64^{+0.08}_{-0.17}$ &  --& -- \\
CEERS-11949      & 214.89190 & 52.93386 &  25.1 & $ 10.65^{+0.29}_{-0.29}$ & $6.52^{+0.34}_{-0.55}$ &  --& --\\
CEERS-57442      & 214.88680 & 52.85538 &  26.0 & $ 10.64^{+0.29}_{-0.31}$ & $7.13^{+0.24}_{-0.36}$ &  --& --\\
CEERS-83784$^*$  & 215.08030 & 52.90790 &  25.5 & $ 10.63^{+0.23}_{-0.29}$ & $7.56^{+0.11}_{-0.15}$ &  --& --\\
CEERS-8825       & 215.01603 & 52.98242 &  23.5 & $ 10.60^{+0.18}_{-0.35}$ & $4.72^{+0.13}_{-0.16}$ &  --& --\\
CEERS-2168       & 215.01127 & 53.01359 &  24.1 & $ 10.59^{+0.27}_{-0.51}$ & $5.11^{+0.19}_{-0.13}$ &  --& --\\
CEERS-87742      & 214.94900 & 52.85175 &  24.6 & $ 10.59^{+0.21}_{-0.24}$ & $4.74^{+0.15}_{-0.22}$ &  --& --\\
CEERS-65868      & 215.08604 & 52.95223 &  23.5 & $ 10.58^{+0.11}_{-0.89}$ & $5.40^{+0.11}_{-0.18}$ & 5.072$^\dag$ & --\\
CEERS-100439     & 214.86580 & 52.77020 &  23.9 & $ 10.52^{+0.21}_{-0.22}$ & $5.27^{+0.14}_{-0.19}$ & --& --\\
CEERS-12787      & 214.85316 & 52.90160 &  23.0 & $ 10.50^{+0.22}_{-0.77}$ & $4.29^{+0.14}_{-0.18}$ & --& --\\
CEERS-68929      & 215.07223 & 52.92533 &  24.6 & $ 10.50^{+0.26}_{-0.37}$ & $5.16^{+0.17}_{-0.17}$ & --& --\\
CEERS-40706      & 214.76108 & 52.75069 &  24.1 & $ 10.50^{+0.27}_{-0.73}$ & $4.00^{+0.18}_{-0.19}$ & --& yes\\
CEERS-26578      & 214.79037 & 52.84192 &  24.0 & $ 10.49^{+0.24}_{-0.19}$ & $4.11^{+0.22}_{-0.24}$ & --& --\\
CEERS-78830      & 214.97479 & 52.86054 &  23.9 & $ 10.49^{+0.25}_{-0.21}$ & $4.05^{+0.25}_{-0.31}$ & --& --\\
CEERS-24868      & 214.83842 & 52.88518 &  24.8 & $ 10.49^{+0.23}_{-0.26}$ & $6.47^{+0.15}_{-0.19}$ & --& --\\
CEERS-95076      & 214.83275 & 52.78136 &  25.4 & $ 10.49^{+0.30}_{-0.23}$ & $4.16^{+0.57}_{-0.36}$ & --& --\\
CEERS-6078       & 215.00289 & 52.98767 &  23.6 & $ 10.47^{+0.15}_{-0.68}$ & $4.75^{+0.14}_{-0.14}$ & --& --\\
CEERS-68303      & 215.08717 & 52.94154 &  23.8 & $ 10.46^{+0.13}_{-0.14}$ & $5.55^{+0.12}_{-0.16}$ & 5.060$^\ddag$ & --\\
CEERS-4851       & 214.97421 & 52.97388 &  23.2 & $ 10.45^{+0.16}_{-0.82}$ & $4.20^{+0.14}_{-0.18}$ & --& --\\
CEERS-14256      & 214.90682 & 52.93192 &  23.0 & $ 10.44^{+0.24}_{-0.76}$ & $4.11^{+0.12}_{-0.20}$ & --& --\\
CEERS-79906      & 214.97836 & 52.85670 &  23.9 & $ 10.44^{+0.24}_{-0.19}$ & $4.64^{+0.14}_{-0.21}$ & --& --\\
CEERS-44057$^*$  & 214.89224 & 52.87741 &  25.9 & $ 10.44^{+0.20}_{-0.27}$ & $7.43^{+0.09}_{-0.13}$ & 6.687$^\ddag$ & --\\
CEERS-84799      & 214.94344 & 52.86410 &  24.1 & $ 10.44^{+0.15}_{-0.09}$ & $4.79^{+0.14}_{-0.17}$ & 4.680$^\ddag$ & yes\\
CEERS-98459      & 214.85289 & 52.77394 &  24.3 & $ 10.43^{+0.25}_{-0.25}$ & $4.80^{+0.21}_{-0.36}$ & --& --\\
CEERS-31016      & 214.80816 & 52.83222 &  24.6 & $ 10.43^{+0.19}_{-0.19}$ & $4.38^{+0.17}_{-0.24}$ & --& --\\
CEERS-78348      & 215.04260 & 52.91196 &  23.6 & $ 10.42^{+0.11}_{-0.10}$ & $4.77^{+0.15}_{-0.14}$ & 4.611\textsuperscript{\ddag}& yes\\
CEERS-73986      & 214.97734 & 52.88961 &  23.5 & $ 10.40^{+0.14}_{-0.78}$ & $4.23^{+0.16}_{-0.18}$ & --& --\\
CEERS-80453      & 215.00767 & 52.87410 &  24.7 & $ 10.40^{+0.24}_{-0.26}$ & $5.98^{+0.27}_{-0.28}$ & --& --\\
CEERS-19333      & 214.94405 & 52.92974 &  23.8 & $ 10.39^{+0.22}_{-0.35}$ & $4.63^{+0.18}_{-0.23}$ & --& --\\
CEERS-75554      & 214.98134 & 52.88257 &  24.5 & $ 10.39^{+0.24}_{-0.23}$ & $4.83^{+0.21}_{-0.29}$ & --& --\\
CEERS-13769      & 214.91046 & 52.93917 &  23.7 & $ 10.38^{+0.16}_{-0.73}$ & $4.41^{+0.14}_{-0.17}$ & --& --\\
CEERS-9235       & 215.03880 & 52.99599 &  23.6 & $ 10.38^{+0.16}_{-0.78}$ & $4.54^{+0.14}_{-0.17}$ & --& --\\
\hline
\hline
\enddata
\tablecomments{Table reporting redshifts and stellar masses of $z>4$ massive galaxy candidates, sorted by mass. Sources with auxiliary data are marked accordingly, the full table can be found in the electronic version of this paper.}
\label{tab:galaxy_properties}
\textsuperscript{*}\footnotesize{EROs} 
\textsuperscript{\dag}\footnotesize{NIRSpec spectroscopic redshift} 
\textsuperscript{\ddag}\footnotesize{NIRCam WFSS spectroscopic redshift} 
\vspace{-8mm}
\end{deluxetable*}

For the majority of galaxies with spectroscopic coverage, we find remarkable agreement between photometric and spectroscopic redshifts, with four sources with $\Delta z > 0.5$ and only one source with $\Delta z > 1$ between the photometric and spectroscopic redshifts (Figure \ref{fig:other_obs}, right panel).  Notably, we find no catastrophic outliers, with all galaxies remaining in our sample with their spectroscopic redshifts, providing confidence in the validity of the full sample of massive galaxies. We show observed spectra for the five NIRSpec prism sources and eight example NIRCam WFSS detections in Figure \ref{fig:sample_spec}, all showing clear emission line detections. We again perform our SED fitting with this updated redshift information.

\begin{figure}
\centering
\includegraphics[width=0.95\linewidth, trim=5mm 5mm 5mm 5mm]{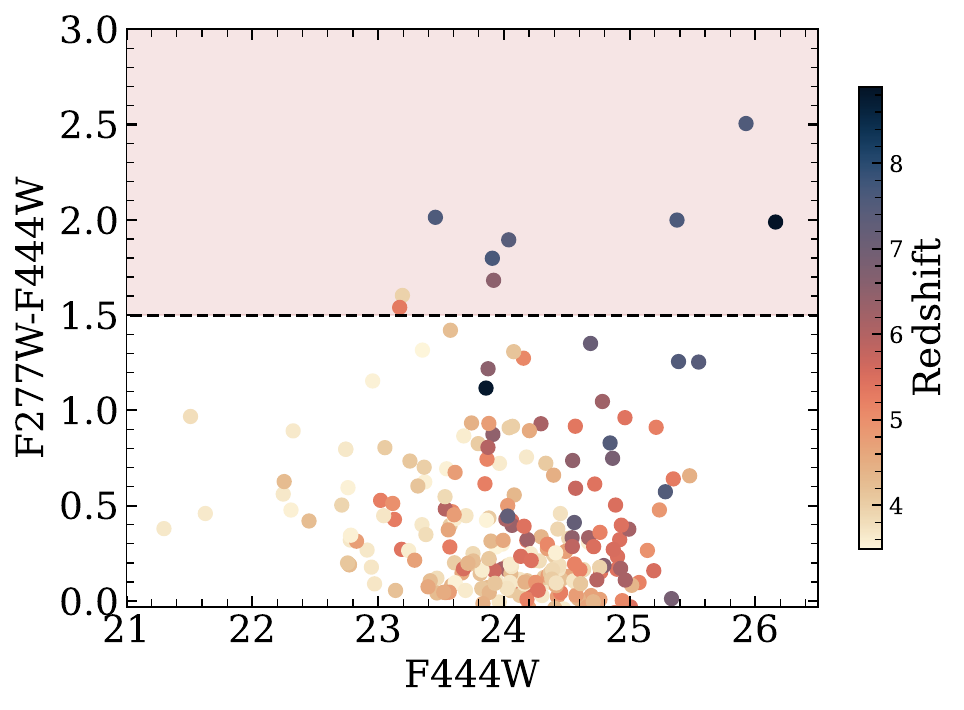}
\caption{Color selection of ``extremely red objects" (EROs) \citep{Barro2023} colored by redshift. These sources are bright and point-source like in F444W, exhibiting blue rest-UV colors with redder rest-optical colors. Due to the brightness of these sources at $\lambda_{observed} > 3 \mu$m, SED fits generally recover large stellar masses. However, these galaxies have photometric colors consistent (and degenerate) with different  combinations of dust-obscured massive star-forming galaxies, less massive non-dust-obscured galaxies, and dust reddened AGNs. Due to these degeneracies, the estimated stellar masses can vary significantly, up to $\sim$3 dex, dependent on assumed galaxy components. Therefore, we remove these sources from our fiducial analysis; however, we still show cumulative number densities found inclusive of these objects for comparison (Figure \ref{fig:number_densities}).}
\label{fig:ERO_colorcut}
\end{figure}

\subsection{Extremely Red Objects}\label{subsec:AGN}
One of the early \textit{JWST} discoveries is the prominence of a new class of sources: high redshift point-like sources that are blue or flat in the short wavelength channels in \textit{JWST} but show extreme reddening towards longer wavelengths. 
\citet{Barro2023} analyzed these sources and found that the most likely explanation of the photometric colors are various combinations of dust-reddened AGN with bluer stellar components. Indeed, early \textit{JWST} observations seem to indicate that accreting super-massive black holes are relatively common at $z>5$ \citep[e.g.][]{Kocevski2023, Larson2023, Leung2023, Harikane2023, Labbe2023UNCOVER, Matthee2023, Furtak2023, Juodzbalis2023}.  This interpretation is supported by a number of such sources with confirmed broad-line AGN \citep{Kocevski2023,Greene2023}, though weaker than expected long-wavelength MIRI fluxes may indicate this population also includes non-AGNs \citep{Williams2023}.

With only photometric colors available, it is extremely difficult to accurately determine the light contributed by the AGN component of these galaxies, making photometric stellar mass estimates for these sources extremely uncertain \citep{Barro2023, Kocevski2023}. These degeneracies mean that estimated stellar masses may vary significantly, up to $\sim 3$ dex dependent on assumed galaxy components \citep{Barro2023}. These uncertainties in stellar mass would bias our results in determining the cumulative number densities of massive galaxies observed; therefore we remove these sources from our fiducial sample with a straightforward color cut of F277W-F444W$>1.5$, which is found to effectively identify sources dominated by potential AGN emission \citep{Barro2023}. 
We show the color-space of our sources in Figure \ref{fig:ERO_colorcut}, as well as the proposed color-cut to remove these objects. We find nine sources which satisfy this color cut, many of which are fit as higher redshift sources ($z>7$). We further visually inspect these and find that they are indeed point-like in F444W.
With this in mind, we perform the following analysis on two samples, one inclusive of all sources determined to have stellar masses of $>10^{10} M_\odot$, and the same sample, but with the red sources described here removed. 

\begin{figure*}[t]
\centering
\includegraphics[width=0.95\textwidth]{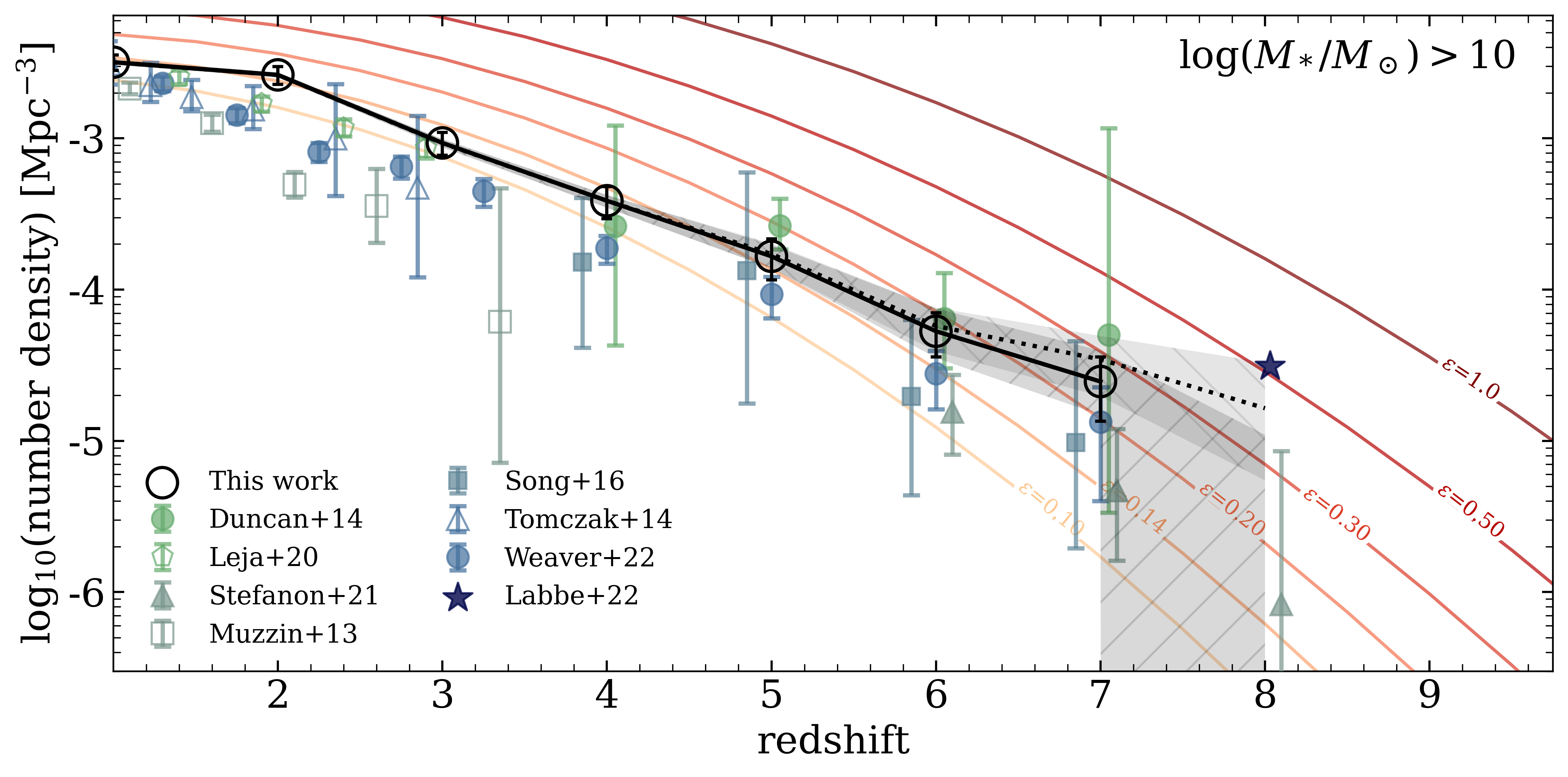}
\caption{The redshift evolution of massive (\logM$>10$) galaxies in CEERS. The black solid (dotted) line shows the volume density for \logM$>10$ galaxies exclusive (inclusive) of EROs, with the darker (lighter) shaded region showing the 68\% confidence interval (CI) inclusive of sampling the redshift and stellar mass posteriors as well as the Poisson error, the errorbars show the error from cosmic variance. We show a variety of pre-\textit{JWST} results (green-blue points) and comparison to the massive galaxies presented in \citet{Labbe2023}, all converted to a Chabrier IMF. The orange-red curves show the expected cumulative number densities above our mass threshold for a given baryon conversion efficiency ($\epsilon$) calculated using  the \citet{Rodrguez-Puebla2016} halo mass function. Our results are systematically higher than previous studies though often within uncertainty, while following a similar evolution to higher redshifts (details in \S \ref{fig:number_densities}) We find that the volume density of massive galaxies follow that expected given a constant baryon conversion efficiency of $\epsilon \sim 0.14$ up to $z\sim5$, before which the volume densities begin to exceed predicted values. 
}
\label{fig:number_densities}
\end{figure*}

In summation, we use SED fitting to identify galaxies with \logM $>10$ at $z = 1-9$ in the CEERS survey. We visually inspect sources in the high redshift sample which are all required to have median $z_{\mathrm{phot}} > 3.5$ and P($z > 2.5$) $> 70\%$; 28 of these galaxies have spectroscopic redshifts. We also identify 9 extremely reddened objects via their F277W-F444W colors and explore the implications of the following results both with and without EROs included in the sample.

\section{Results}\label{sec:results}
\subsection{Cumulative Number Densities of Massive Galaxies} \label{subsec:number_densities}

To determine the cumulative number densities of massive galaxies across cosmic time, we separate our sample into redshift bins of $\Delta z = 1$. At $z<4$, we take the median fitted physical parameters for each source recovered from \texttt{dense basis}. Above $z>4$ we draw from the recovered posterior distribution for each individual (visually inspected) source. For each source, we conduct a draw from the posterior SED to obtain a stellar mass and redshift (stellar mass and redshift sampled as a pair as they are co-variant), and assign the galaxy to a redshift bin if the drawn stellar mass is greater than our intended mass limit (\logM $> 10$). We perform this sampling over 500 draws to recover the number of massive galaxies across redshift bins. For each draw, we further sample from the Poisson distribution of the number of galaxies in each redshift bin. Combined, these provides uncertainties in the number densities inclusive of the spread in redshift, stellar mass, and Poisson uncertainties. To account for cosmic variance, or the variance that arises from the effects of large-scale structure, we use the approach presented in \citet{Yung2023RomanLightCone} and calculate the variance from 2500 random samples of a 6'$\times$14.7' elongated field from 5 of the 2-deg$^2$ lightcones presented in that work. In Figure \ref{fig:number_densities}, we report the median as the number densities and show the 68\% confidence limit as the shaded region, the CV is shown as error bars. 

We show in Figure \ref{fig:number_densities} the cumulative number densities of \logM$>10$ galaxies as recovered from this analysis in both our entire galaxy sample, and the same sample with red point-sources removed (\S \ref{subsec:AGN}) in the CEERS field, as a function of redshift. For comparison, we also plot curves of cumulative number densities of \logM $>10$ galaxies given various values of constant baryon conversion efficiency, or the rate of conversion from baryonic to stellar mass using halo mass functions from the Bolshoi-Planck and MultiDark-Planck $\Lambda$CDM cosmological simulations \citep{Rodrguez-Puebla2016}.
Assuming a cosmic baryon fraction $f_{\text {baryon}}$ = 0.16, which is consistent with the value established by observations of the cosmic microwave background \citep{Spergel2003, Jarosik_2011}, we convert the underlying halo mass function to a stellar mass function. We then integrate the converted stellar mass function of \logM $>$ 10 sources assuming varying values of efficiency, $\epsilon$. 

In CEERS, we see that above a redshift of $z\sim4$, while the number density of massive galaxies continues to decrease with increasing redshift, there is an increased abundance over that expected with a constant baryon conversion efficiency as we progress to higher redshifts. This increased abundance could be explained by a difference in star-formation physics in the early Universe, resulting in an increased global baryon conversion efficiency and a non-standard galaxy initial mass function, discussed in \S 5.

\subsection{Sizes of Massive Galaxies} \label{subsec:sizes}

\begin{figure}[t]
\centering
\includegraphics[width=0.95\linewidth, trim=5mm 5mm 5mm 5mm]{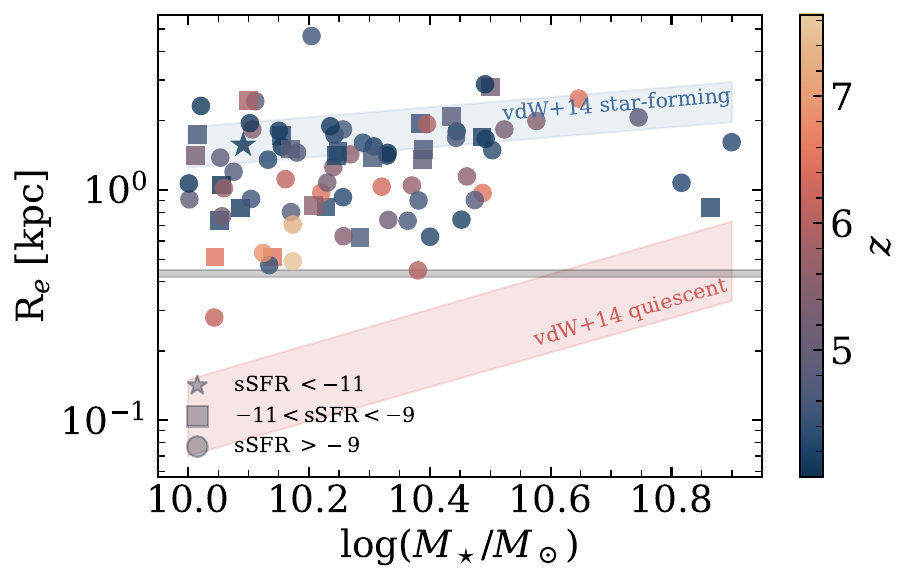}
\caption{The size-mass distribution of $z>4$ massive galaxies presented here. We show the effective semi-major axes measured at rest-frame 0.5 $\mu$m. The points are colored by redshift and the shape indicates the 10Myr averaged sSFR values from \texttt{dense-basis}. The best-fit size-mass relations for quiescent and star-forming galaxies from \citet{vdWel2014} extrapolated to $z=4-8$ is also shown for comparison. The grey line shows the size range of a point source from $z=4-7$. We find that our sample is primarily star-forming and the slope of our sample in the size-mass plane is consistent with results for star-forming galaxies at lower redshifts, however, they are smaller by a factor of $\sim 2$ than predicted when the relation is extrapolated out to higher redshifts.}
\label{fig:size-mass}
\end{figure}

In Figure \ref{fig:size-mass}, we present the size-mass distribution of our massive sample from $4<z<8$. Sizes were measured with {\sc galfit} and are effective semi-major axes measured at rest-frame 0.5 $\mu$m. Details of the {\sc galfit} fitting will be provided in a future paper (McGrath et al., in prep.).  Briefly, sources were fit using empirical PSFs \citep{Finkelstein2023b}, and the ERR array was used as the input sigma image, which includes noise elements from the detector, sky, and source. Neighboring galaxies were either fit simultaneously or masked during fitting, depending on brightness and distance from the primary source. Only galaxies whose best-fit models yielded magnitudes consistent with our photometry and which did not reach a constraint limit are shown (83/118 sources). We also show the best-fit size-mass relations from \citet{vdWel2014} extrapolated to this redshift range for comparison.

The slope in the size-mass plane for our massive galaxies sample is consistent with results for star-forming galaxies at lower redshift, but when we extrapolate the evolving intercept of the size-mass relation from lower redshift to $z >$ 4, we find that galaxies at $4<z<8$ are smaller on average by a factor of $1.5-2$ than predicted from previous work \citep{vdWel2014}. This is consistent with the results of \citet{Ormerod2023} and \cite{Ward2023}, who also found evolution towards smaller sizes at higher redshifts. Furthermore, we see evidence for continued redshift evolution, with $z=6-7$ galaxies smaller at fixed mass than $z=4-5$ galaxies by a factor of 1.67, implying that the processes responsible for driving this evolution are already in place at very early times in the universe.

\subsection{Formation Timescales} \label{subsec:SFHs}

Star-formation histories, or the SFR of a galaxy across its lifetime, probes galaxy formation and evolution. By studying the star-formation histories of massive galaxies in particular, we can attempt to understand how these galaxies manage to form surprisingly large amounts of stellar mass over relatively short timescales, and provide valuable insights into the early stages of galaxy formation and crucial benchmarks for galaxy formation theory.

\begin{figure}[t]
\centering
\includegraphics[width=0.95\linewidth, trim=5mm 5mm 5mm 5mm]{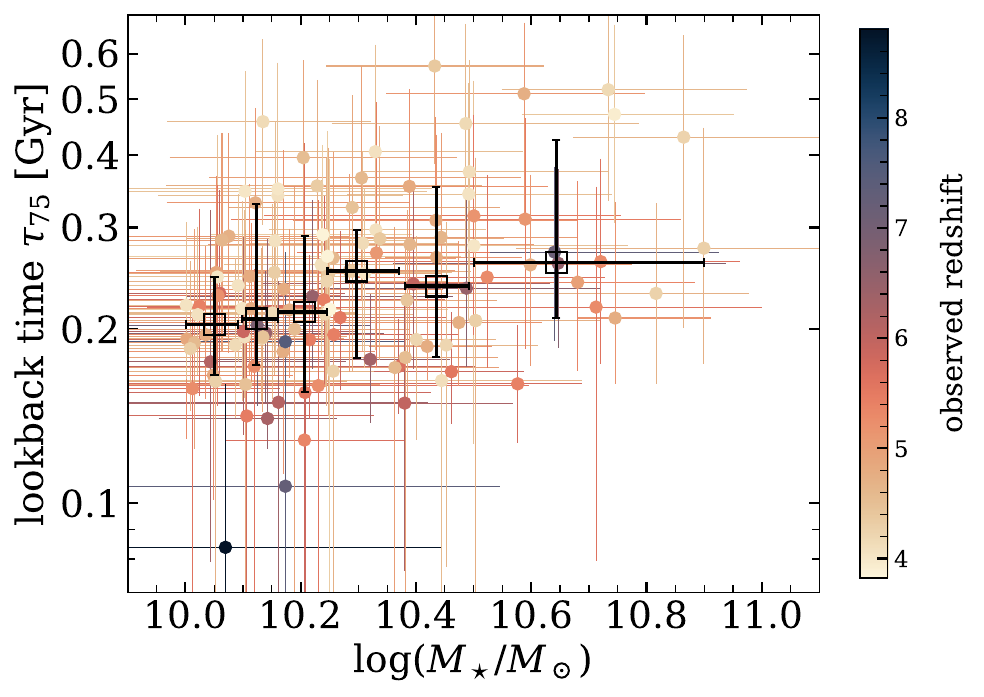}
\caption{The estimated formation time of 75\% of the stellar mass (in terms of lookback time, $\tau_{75}$) for each galaxy in our massive, high redshift sample. The points are colored by redshift, we bin galaxies into bins of stellar mass, each bin with an equal number of sources; the open black squares show the median of each bin, with the error bars denoting the 18\% and 84\% spread. There is a slight trend with $\tau_{75}$ with respect to the observed stellar mass, with galaxies of larger mass forming the bulk of their stars at an earlier time, but this is not significant with constraints on these quantities from the current data. 
Due to the difficulty constraining star-formation histories with only photometry, we note there exist large errors associated with $\tau_{75}$. SFHs are often difficult to constrain based on photometry alone, with follow-up spectroscopy often necessary to properly disentangle the underlying stellar populations. Furthermore, at very high redshifts, our sample is small ($\sim$ a few galaxies at $z>7$); larger surveys will be able to provide the sample sizes needed to properly infer trends with redshift. }
\label{fig:SFH_spread}
\end{figure}

From \texttt{dense basis}, we estimate the times at which each galaxy formed 25\%, 50\%, 75\% of its total stellar mass at the time of observation and the associated spread in each timescale. Figure \ref{fig:SFH_spread} shows the distribution of $\tau_{75}$ (the difference between the time of observation compared to the time when the galaxies formed 75\% of their stellar mass) for our sample with median \logM $>10$ with EROs removed (120 sources).
We separate our sample into 5 bins of equal number of galaxies across stellar mass and show the median $\tau_{75}$ and 68\% spread in each bin. We notice that there may be a slight dependence on $\tau_{75}$ at which 75\% of the observed stellar mass formed with respect to current observed stellar mass, similar to results found by \citet{Estrada-Carpenter2020}; 
however the spread and errors are large, making this relationship difficult to robustly constrain. 

We show some example fitted SFHs in Figure \ref{fig:SFH_plots}. With \texttt{dense basis}, we recover a diversity of SFH shapes, including periods of bursts, and rising/falling SFHs. We note that the posterior SFHs often have a large spread in recovered shapes, with some sources exhibiting unconstrained SFHs, and others which are more constrained. 
Without spectral information, it is difficult to disentangle the emission/absorption lines from the continuum for each galaxy; therefore the exact stellar populations also become difficult to fully constrain,
particularly for older stellar populations which are often out-shined by younger more massive stars \citep{Papovich2023, Conroy2013}. While recent bursts of star-formation activity can be indicated by photometry, spectroscopic observations are necessary to robustly constrain the full mass assembly histories of these galaxies \citep{Estrada-Carpenter2020, Estrada-Carpenter2021}. Furthermore, larger samples of massive galaxies
may be able to recover more robust trends with star-formation timescales.

\begin{figure*}
    \centering
    \input{SFH_plots.tex}
\end{figure*}

\section{Discussion}\label{sec:Discussion}
\subsection{Comparison to Literature Results} \label{subsec:comp_obs}

Here we compare our measured cumulative number densities to those inferred from previously published stellar mass functions, converting all results to a Chabrier IMF where needed (Figure \ref{fig:number_densities}). At lower redshifts, ($z<4$), we compare to results from 
\citet{Tomczak2014}, who surveyed an area of 316 arcmin$^2$ distributed over three independent fields, using observations from the FourStar Galaxy Evolution Survey (ZFOURGE). Our results show slightly higher number densities that agree within the quoted uncertainties. We also show results from the $K_s$ selected catalog of the 1.62 deg$^2$ COSMOS/UltraVISTA field  from $0.2<z<4.0$  from \citet{Muzzin2013}. 
Our results show higher number densities; however, we note that  beyond $z \geq 2.0$, the analysis from \citet{Muzzin2013} is limited to galaxies with \logM $>10.51$, reducing its measured number densities. We also show recent work from \citet{Leja2020}, who compute an updated SMF from SED fitting with non-parametric SFHs of $\sim 10^5$ galaxies in the 3D-HST and COSMOS-2015 surveys to infer updated stellar masses, and constructs a continuity model that directly fits for the redshift evolution of the SMF. 

At higher redshifts, our results are overall higher than the stellar mass functions from Great Observatories Origins Deep Survey Re-ionization Era Wide-area Treasury from Spitzer program \citep[GREATS:][]{Stefanon2021}. We are within the error of the stellar mass function presented by \citet{Song2016}, calculated from a rest-frame UV selected sample of $\sim$4500 galaxies, over the $\sim$280 arcmin$^2$ CANDELS/GOODS HUDF fields, and \citet{Duncan2014} who measure the SMF in the CANDELS GOODS-South field from $4<z<7$. 
\citet{Weaver2022} studied the galaxy SMF from redshifts $0.2<z\leq 7.5$, leveraging the 1.27 deg$^2$ coverage of the COSMOS2020 catalog \citep{Weaver2020COSMOS_catalog}. We find overall higher number densities across all redshifts, while following the same general trend as we push to higher redshifts.

While resulting cumulative number densities in this work are generally higher than previous works, our results make use of deeper data covering both the rest-frame UV and optical at these redshifts. The improvement in angular resolution compared to \textit{Spitzer/IRAC} largely corrects the source blending seen with \textit{IRAC}, resulting in higher completeness.
We note that in particular, the number density inclusive of EROs we find here (if extrapolated) agrees within the uncertainties with previous studies of massive galaxies in the CEERS field by \citet{Labbe2023}. 

\subsection{Comparisons to Models} \label{subsec:comp_models}

\begin{figure}[t!]
\centering
\includegraphics[width=1.0\linewidth]{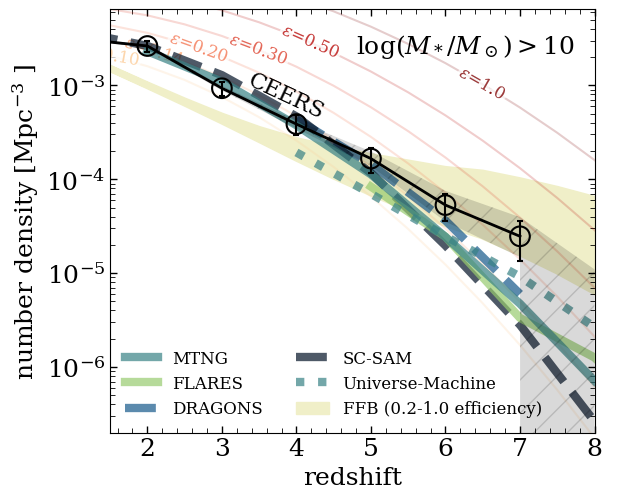}
\caption{Comparison of the number density of \logM $>10.$ galaxies found in CEERS vs. models. Here we show the sampled number densities with EROs removed (details in \S \ref{subsec:AGN}), the grey shaded region in our 95\% CI. We note that at lower redshifts ($z<4$), models seem to agree well; however, they diverge at higher redshifts. Overall, we find good agreement with DRAGONS, MTNG, and SC-SAM at $z < 4.5$. At $z>5$ we find higher median number density than predicted by all models shown aside from FFB, where our results are best fit by the FFB predictions with $\epsilon_{\rm max} = 0.3$ at $z>6$.
}
\label{fig:model_comp}
\end{figure}

We show in Figure \ref{fig:model_comp} comparisons between the number densities of massive galaxies in CEERS and a variety of models, including hydrodynamical simulations Millennium-TNG \citep[MTNG,][]{Kannan_2023, Kannan2022MNRAS.511.4005K,Springel2018MNRAS.475..676S} and FLARES \citep{Lovell2021FLARES, Wilkins2023FLARES}, the empirical model Universe-Machine \citep{Universe-Machine}, and semi-analytic models DRAGONS \citep{DRAGONSsim} and the Santa Cruz SAM \citep[SC-SAM,][]{Somerville2015, Yung2019a, Yung2019b}, and the Feedback-Free Starbursts \citep[FFB,][]{Dekel2023} model.

While the models agree well at low redshift, they tend to diverge at higher redshifts due to a combination of effects from physical assumptions and the volume probed. 
Below a redshift of $z \sim 5$ (i.e. $z < 5$), our results agree well with SC-SAM, MTNG, and DRAGONS, however we find higher abundances of massive galaxies than that predicted by FFB, which is made by construction to converge with the Universe-Machine at low redshifts. At $z \gtrsim 5$, we find a higher median cumulative number density than almost all of the models that we compared with except for FFB, which argues for high star-formation efficiency due to FFB in massive galaxies at cosmic dawn, where the gas density in star-forming clouds is above a critical value \citep{Dekel2023}. The shaded area corresponds to a range of maximum efficiency from $\epsilon_{\rm max} = 0.2$ (bottom) to $1.0$ (top). Our results at $z>6$ most closely follow the FFB model with $\epsilon_{\rm max} \sim 0.3$.

\subsection{Evolution of Massive Galaxies}\label{subsec:evolution}

\textit{JWST} is revealing a high abundance of bright galaxies in the early Universe \citep[e.g.][]{Finkelstein2023, Castellano2022, Casey2023, Franco2023, Leung2023, Naidu2022, Adams2023arXiv230413721A, Bouwens2023, Harikane2023}.
These observations may be explained by changing the physical processes that govern the relationship between observed galaxy light and the host dark matter halo mass. 
 In this analysis we find a similar trend of a high abundance of massive galaxies in the early Universe. Here, we discuss two possible physical processes that can explain this excess: 1. a higher baryon conversion efficiency at high redshifts, and 2. a changing mass-to-light ratio, where the assumptions made in converting the observed light to stellar mass grounded in the local Universe no longer hold at high redshifts. Other physical processes may also be invoked to explain excesses in UV luminous and massive galaxies, such as modifying cosmology, or varying dust attenuation and/or levels of stochasticity of star formation, though we note that the latter two are encompassed in the performed SED modelling and should not effect the recovered stellar masses significantly; we refer the reader to works by \citet{Mason2023, Finkelstein2023b, Pallottini2023, Shen2023, Sun2023BurstySFH, Ferrara2023, Dressler2023, MBK2023, Parashari2023, Maio2023} for further discussion on these particular topics.

\subsubsection{A Redshift Dependence in Global Baryon Conversion Efficiency} \label{subsubsec:SFefficiency}
The baryon conversion efficiency ($\epsilon$) describes how efficiently the baryonic gas accreted onto dark matter halos is converted into stars. It is defined as:
\begin{equation}
    M_\star = \epsilon f_\mathrm{baryon}M_\mathrm{halo}
    \label{eq:efficiency}
\end{equation}

This efficiency, as defined here, is averaged over the star-formation history of the galaxy, and is determined to strongly correlate with both halo mass as well as the stellar mass of the central galaxy \citep{Moster2010, Behroozi2010}.
In the local Universe, low values of $\epsilon$ ($\lesssim 0.15$) are inferred \citep{Bregman2007, Zhang2022}. 
Energetic feedback from a combination of effects including stellar winds, supernovae, AGN, mergers/shock heating and morphological quenching are thought to quench and prevent large amounts of efficient star-formation from occurring in the local Universe \citep[e.g.][etc.]{Kauffmann_Charlot_1998, Grimes2009, Kondapally2023, Martig2009, Somerville_Dave_2015, NaabOstriker2017}.
\citet{Dekel2023} have argued that at extremely high redshifts ($z \gtrsim 10$), $\epsilon$ may approach unity because in the early Universe,
the combination of high ISM density and low metallicities implies that molecular cloud freefall times are shorter than the time it takes for stars to develop winds and supernovae. 
When they explode in higher density gas, supernovae explosions become less efficient at heating and driving winds, perhaps causing weaker stellar feedback in the early universe \citep{Walch2015}.

\begin{figure}[t]
\centering
\includegraphics[width=0.9\linewidth]{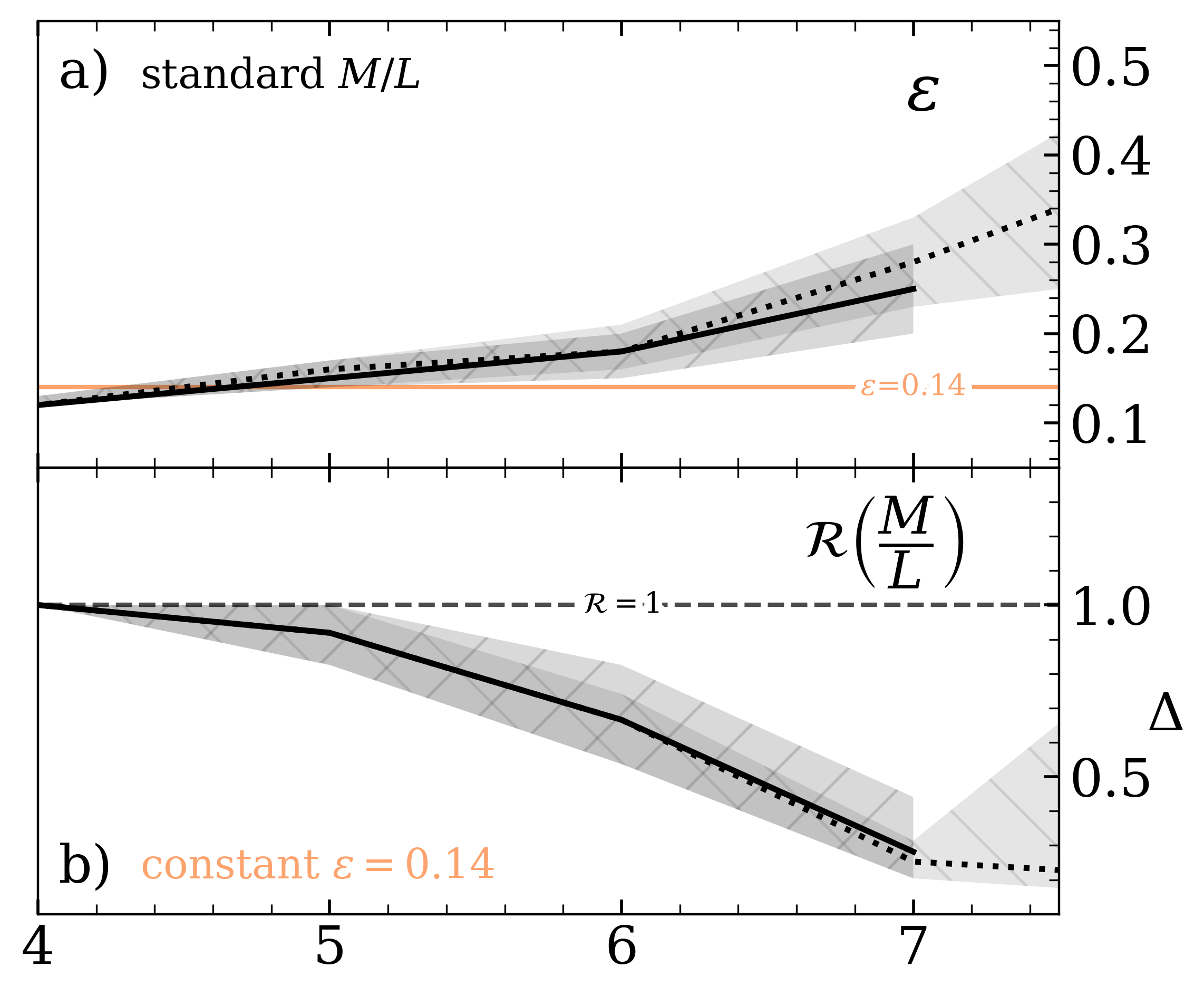}
\caption{
\textit{a)} The evolution in baryon conversion efficiency given a constant mass-to-light ratio needed to match our observations of massive galaxies (\S \ref{subsubsec:SFefficiency}). Or, \textit{b)} the factor change in mass-to-light required to match the number densities of \logM $>10$ expected assuming a constant efficiency of $\epsilon = 0.14$. We find that the observed cumulative number densities in CEERS follow that expected from a constant baryon conversion efficiency of $\sim 0.14$ up to a redshift of $z \sim 4$, before which we find an excess of \logM $>10$ galaxies, potentially indicating that the global baryon conversion efficiency is higher \emph{or} that the mass-to-light ratio is lower in the early Universe than at lower redshift.
}
\label{fig:number_den_changes}
\end{figure}

To explore how the baryon conversion efficiency may have had to evolve in order to match the observed number densities for each redshift bin, we find the cumulative number density for each value of $\epsilon$ that most closely matches the observed cumulative number density in CEERS. From this, we show how the baryon conversion efficiency would evolve with redshift to explain the surplus of massive galaxies in the early Universe. 
We find that, to match the observed number of \logM $>10$ galaxies in this study, the efficiency remains relatively constant ($\epsilon \sim 0.14$) from $z =$ 1 to $z \sim 4$, at which point it steadily increases, reaching$\epsilon \sim 0.3$ at $z \sim 7$, a factor of $3\times$ more efficient than the average efficiency expected in the local Universe (\citet{Bregman2007, Zhang2022}; Figure \ref{fig:number_den_changes}; panel a).

 If the light observed from EROs is indeed primarily contributed by their stellar components, and thus the estimated masses are robust, we show that there would need to be a dramatic increase in $\epsilon$, particularly at $z > 6$, where they represent a significant portion of massive galaxies in our sample. Including these sources the baryon conversion efficiency would have to have reached a value of $0.4$ by $\sim 7$. 
 At even higher redshifts \citet{Dekel2023} predicts an even higher efficiency value (close to unity at $z \sim 10$). While we do not probe to such high redshifts in this work due to our mass limit combined with the cosmic volume probed (the area coverage of CEERS is not expected to find rare, extremely massive high redshift sources \citealt{Santini_2021}).
 
Our inference of $\epsilon$ as defined in Eq.~\ref{eq:efficiency} requires knowledge of the underlying dark matter halo mass function. We use mass functions obtained directly from large cosmological simulations (Bolshoi-Planck and MultiDark-Planck) run with cosmological parameters that are consistent with the Planck cosmology (see Sec.~\ref{subsec:number_densities}). If the true halo mass function deviates from these simulation results, the inferred star formation efficiencies will change accordingly. However, this would only affect our results at a quantitative rather than qualitative level unless the background cosmology itself deviates from the flat Planck cosmology. For a recent detailed investigation of halo mass functions at high redshift, see \citet{Yung2023}. 

\subsubsection{Changing Mass-to-Light Ratio} \label{subsubsec:MtoL}

Another explanation for the implied excess of massive galaxies may be due to an incorrect assumption about the mass-to-light ratio at high redshifts. The determination of the stellar mass function (SMF) is a cornerstone in astrophysics, for the stellar mass distribution determines the evolution, surface brightness, chemical enrichment, and baryonic content of galaxies. SED fitting codes estimate the mass of the galaxy from its observed brightness, dependent on an assumed initial mass function (IMF), or the assumption of the abundance of stars at any given mass populating the galaxy. 
Various functional forms of the IMF are widely used \citep[e.g.][]{Salpeter1955, ScaloMiller1979, Kroupa2002, ChabrierIMF}. The determination of the initial mass function is not direct; but instead is dependent on the transformation of the observed light from objects into mass reliant on theories of stellar evolution. Current IMFs are calibrated on a Galactic scale, where individual stars are resolvable, while the IMF characteristics of early star formation at large redshift remain undetermined. However, changes in the characteristic stellar mass for an IMF may be expected in the early Universe \citep{Steinhardt2023, Sneppen2022}. The IMF depends upon several properties of the star-forming clouds, with the high temperature and low gas metallicities in the early Universe motivating the formation of more massive stars. This can lead to the decrease in mass-to-light ratio by a factor of several \citep[e.g.][]{Bromm1999, Raiter2010}, which could also explain the high observed abundances of UV bright galaxies at $z>10$ from recent \textit{JWST} observations \citep{Finkelstein2023, Harikane2023, Yung2023}.

In Figure \ref{fig:number_den_changes}, we show how the mass measured from our SED fitting would need to be adjusted in order for the cumulative number densities of massive galaxies observed to match those expected from a constant baryon conversion efficiency ($\epsilon=0.14$). Our reported stellar masses are determined using a Chabrier (2003) IMF, thus the change in the mass-to-light ratio shown here is the deviation from such. To determine this, we repeat the procedure as detailed in \S \ref{subsec:number_densities}, drawing from the posteriors of the SED fits, and bin them according to mass and redshift. For each redshift bin, we then divide all masses by some value $\delta$  (starting at 1.0 with increments of 0.0001) until the number of galaxies with \logM $> 10$ is within 10\% of that which is expected given an assumed constant baryon conversion efficiency of $\epsilon = 0.14$ (Figure \ref{fig:number_den_changes}; panel b). This analysis shows that the mass-to-light ratio should be $\sim 2 \times$ lower by $z=6.5$ and $\sim 3 \times$ lower by $z=7$. These ratios are consistent with those expected of top-heavy IMFs, results from \citet{Yung2023, Yung2023GUREFT} semi-analytic model predictions based on the GUREFT simulation suite shows that a $3 \times$ decrease in the mass-to-light ratio can be invoked to match the excess of UV luminous galaxies and match the observed UVLF at $z \sim$ 11.

\section{Conclusion}\label{sec:conclusion}

In this work, we study the evolution of massive (\logM $>10$) galaxies in the CEERS survey. The launch of \textit{JWST} has brought with it an unparalleled ability to measure stellar masses at redshifts beyond $z>5$, which was historically limited due to the abilities of previous telescope facilities to constrain rest-frame optical emission. 

To select massive galaxies, we perform SED fitting with \texttt{eazy} to obtain photometric redshift probabilities and with \texttt{dense basis} to fit flexible non-parametric SFHs and recover posteriors on stellar masses. We perform visual inspection of all high redshift ($z>3.5$) sources for which the 97$^\textrm{th}$ percentile posterior in stellar mass is \logM $>10$ (561 sources), allowing us to further vet photometry and their resulting fits. We remove a total of 300 sources from our final high redshift sample and note that this significant number of contaminant sources speaks to the necessity of careful selection for these rare galaxies. We remove likely AGN contaminants from our sample, as the stellar mass estimates cannot be accurately determined without further information on light contribution from the central AGN. We then sample the posterior fit for each galaxy and take the median $\pm 1\sigma$ spread and arrive at a cumulative number density inclusive of errors from SED fitting, Poisson, and cosmic variance. From this analysis, we find that the volume densities of massive galaxies in CEERS follow that expected from a constant baryon conversion ratio of $\epsilon \sim 0.14$ up to $z \sim 5$, before which the volume densities exceed a constant $\epsilon$ and appear to indicate higher $\epsilon$ at higher redshifts.

We find that the number of $z>4$ massive galaxies in CEERS exceeds most theoretical predictions, particularly at redshifts above $z \sim 5$. While we do find a higher than expected cumulative number density at $z>4$, our findings are entirely consistent with the $\Lambda$CDM cosmological model if we posit an increased efficiency of conversion of baryons into stars in massive dark matter halos with increasing redshift.
We show that a redshift-dependent global baryon conversion efficiency or a change to the mass-to-light ratios of galaxies' stellar populations, e.g., due to the variation in the IMF can both reproduce the number of observed massive galaxies at high redshifts. 
To disentangle these scenarios, further observations to measure galaxy clustering may provide insights into star-formation efficiencies \citep{Munoz_2023} while deep spectroscopic observations to detect very massive stars via strong hydrogen and helium recombination lines alongside a lack of heavy metal features to provide constraints on assumed IMFs \citep{Trinca2023, Yung2023, Venditti2023}.

\begin{acknowledgments}
K.C. and S.L.F. acknowledge support from the National Science Foundation through grant AST-2009905 and from NASA through STScI award JWST-ERS-1345. This material is based upon work supported by the National Science Foundation Graduate Research Fellowship under Grant No. DGE 2137420. MBK acknowledges support from NSF CAREER award AST-1752913, NSF grants AST-1910346 and AST-2108962, NASA grant 80NSSC22K0827, and HST-AR-15809, HST-GO-15658, HST-GO-15901, HST-GO-15902, HST-AR-16159, HST-GO-16226, HST-GO-16686, HST-AR-17028, and HST-AR-17043 from the Space Telescope Science Institute, which is operated by AURA, Inc., under NASA contract NAS5-26555. 
\end{acknowledgments}

%

\vspace{5mm}
\facilities{HST(STIS), \textit{JWST}(STSci)}


\software{astropy \citep{2013A&A...558A..33A,2018AJ....156..123A},  EAZY \citep{Brammer2008},
Source Extractor \citep{1996A&AS..117..393B}, 
Dense-Basis \citep{Iyer_2019, Iyer_2017},
Colormaps from \citet{crameri_fabio_2023_8035877}}




\bibliography{CEERS_massive}{}
\bibliographystyle{aasjournal}



\end{document}